# A New Web Based Student Annual Review Information System (SARIS) With Student Success Prediction

Adeel Akbar Memon[1], Chengliang Wang[2], Muhammad Rashid Naeem[3],
Muhammad Tahir[4], Muhammad Aamir[5]
*(School of Software Engineering, Chongqing University, P.R. China)*

***ABSTRACT :*** *In this paper, we are proposing new web based Student Annual Review Information System (SARIS) and prediction method for the success of scholar students to China Scholarship Council(CSC). The main objective of developing this system is to save the cost of paper, to reduce the risk of data loss, to decrease the processing time, to reduce the delay in finding for the successful students. The proposed system and prediction method is intended to be used by China Scholarship Council; however SARIS and prediction method are quite generic and can be used by other scholarship agencies.*

***Keywords –*** *China Scholarship Council, database system, data mining, information system, student success.*

## I. INTRODUCTION

Every year, the China Scholarship Council is offering thousands of various types of scholarships including Chinese Government Scholarship, Confucius Institute Scholarship etc. They are offered either full or partial scholarships. These scholarships are offered for two or more years. CSC keeps track of each scholar student every year by asking the student to submit an Annual Review Form. The scholarships are continued based on the academic performance of a student. The approval/disapproval for continuing scholarship is based on few factors identified by CSC [4].

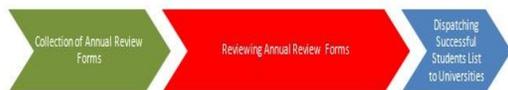

Figure 1.1: Time wise representation of tasks

The latest inventions in science and technology have changed our lives and style of living. Technology has moved almirah of papers into Databases [1,2] and handwriting work to filling forms on the Information Systems [3] using the power of Internet [6]. One of the best solutions for computerizing manual Student Annual Review process is to develop a Management Information System [7].

The topic of this paper focuses on proposing a new web based solution named as Student Annual Review Information System (SARIS) and prediction method for successful students by using data mining [8].

Our paper is organized as follows. Section II is related with the background of the system. In section III, we have analyzed the CSC database. Section IV is related with the SARIS database model. Section V represents the SARIS application privileges and access rights. Section VI deals with the proposed prediction method for predicting the students' success. Finally, we have concluded our research work in Section VII.

## II. BACKGROUND

The Information System is a set of different components used to collect, store, and process data to deliver information or knowledge. Nowadays, different types of information systems are in use. The following figure illustrates the pyramid model for information systems.

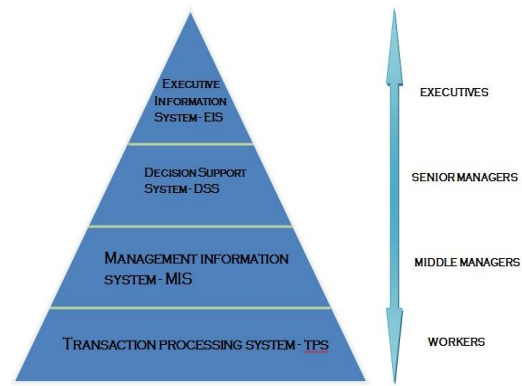

Figure 2.1: Pyramid model for IS

Management Information System – MIS

Management Information System is followed by TPS in the pyramid. MIS provides the information require by organizations to manage themselves efficiently and effectively. There are five primary components: Hardware, Software, Data, Procedures and People. Usually, middle managers are the users of MIS.

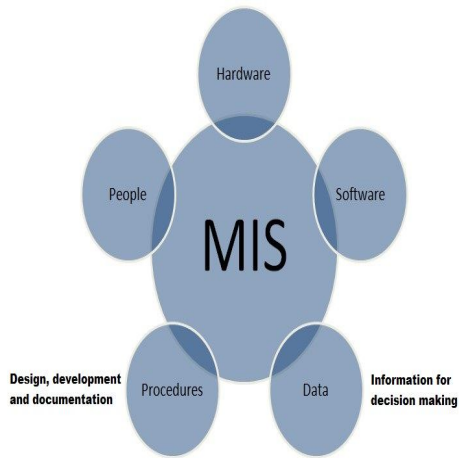

Figure 2.2: Five components of MIS

Data Mining

Data Mining, the important analysis step of Knowledge Discovery in Databases (KDD) is the process of discovering patterns in datasets (usually large data sets). The goal of data mining process is to extract the useful information from data set and transform (using transformation techniques) it into understandable structure.

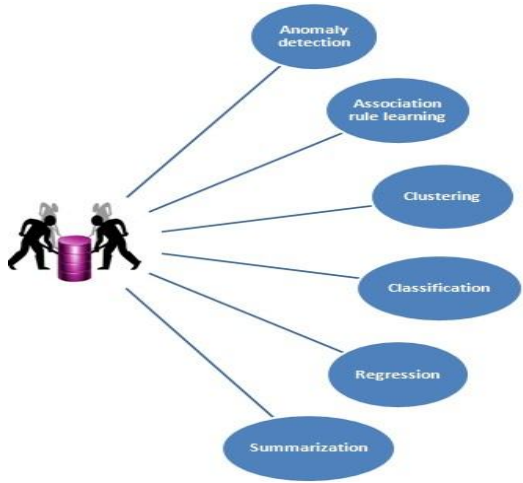

Figure 2.3: Six common classes of tasks

### III. ANALYSIS OF CSC WEB PORTAL

We analyzed China Scholarship Council web portal [5] and identified that the following tables are already used by CSC: STUDENT_T, SCHOLARSHIP_T, SCHOLARSHIP_TYPE_T, MAJOR_FIELDS_T, PERIOD_T, UNIVERSITY and SCHOOL_T. The table and attribute names are taken as an assumption. There may be different tables attributes, and number of attributes, but we have considered the most useful ones into our account.

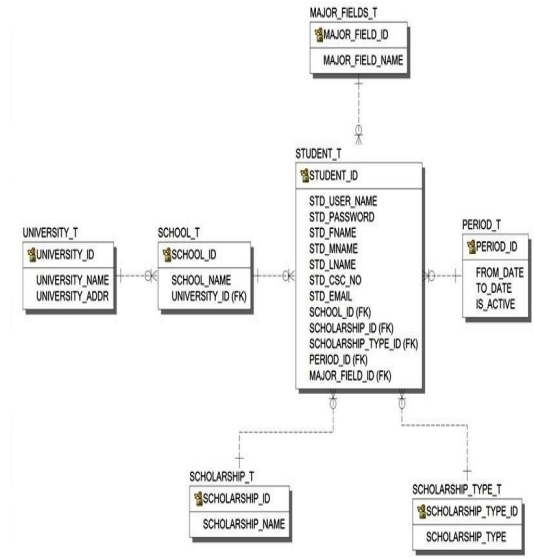

Figure 3.1: CSC Database ER diagram

We identified the use/purpose of each table as under:

1. STUDENT_T: This table contains the registration details, basic information, admission related school/university information, major field of study, scholarship name, period, and scholarship type information of student.
2. SCHOLARSHIP_T: This table contains the names of scholarships offered by China Scholarship Council.
3. SCHOLARSHIP_TYPE_T: This table holds the information about type of scholarship either Full or Partial scholarship.
4. MAJOR_FIELDS_T: This table captures the information for the majors taught in the universities located in China.
5. PERIOD_T: This table contains the information regarding the start and end ate of the scholarship.
6. UNIVERSITY_T: Table holds the names of the universities in China.
7. SCHOOL_T: This table contains the information for schools related with the universities.

### IV. SARIS DATABASE MODEL

In this section a new database model is proposed for SARIS. In contrast to the CSC database, we felt the necessity of the other tables which are used to support and collect Student

Annual Review information. Few of them are supporting in the prediction method for student success.

Figure 4.1: Entity Relationship Diagram for SARIS
The purpose of tables used is described as under:

1. STUDENT_AREVIEW_T: The most important table of the SARIS database. The table contains all the information related to the Student Annual Review Form including short summary entered by student, academic score, punishments, rewards, and review summary entered and verified by the reviewer.
2. REVIEWER_T: The supportive table for Student Annual Review and required for authenticating and verifying the student's short summary. The table contains the basic and employee information of the reviewer.
3. TEACHERS_T: This table is not in use for the SARIS application. This table is only used to match the reviewer information so that fake reviewer accounts can't be created. This table also holds the basic and employee information of the teachers of schools of universities.
4. SUBJECTS_T: The table contains the information about the subjects taught in any major of school including the total number of marks and the total no. of hours for this subject.
5. STUDENT_SCORE_T: Table holds the information about the student's academic score and the no. of hours attended.
6. STUDENT_PUNISHMENTS_T: The purpose of table is to hold the information regarding the punishments including the seriousness of punishments.
7. PUNISH_SERIOUSNESS_T : The table is used to contain the information regarding the seriousness of the punishments such as "Warning", "Serious Warning1" and "Serious Warning2". It also contains the information that either this punishment can cause the dismissal from scholarship or not.
8. STUDENT_REWARDS_T: Table holds the information regarding the rewards awarded to the student.

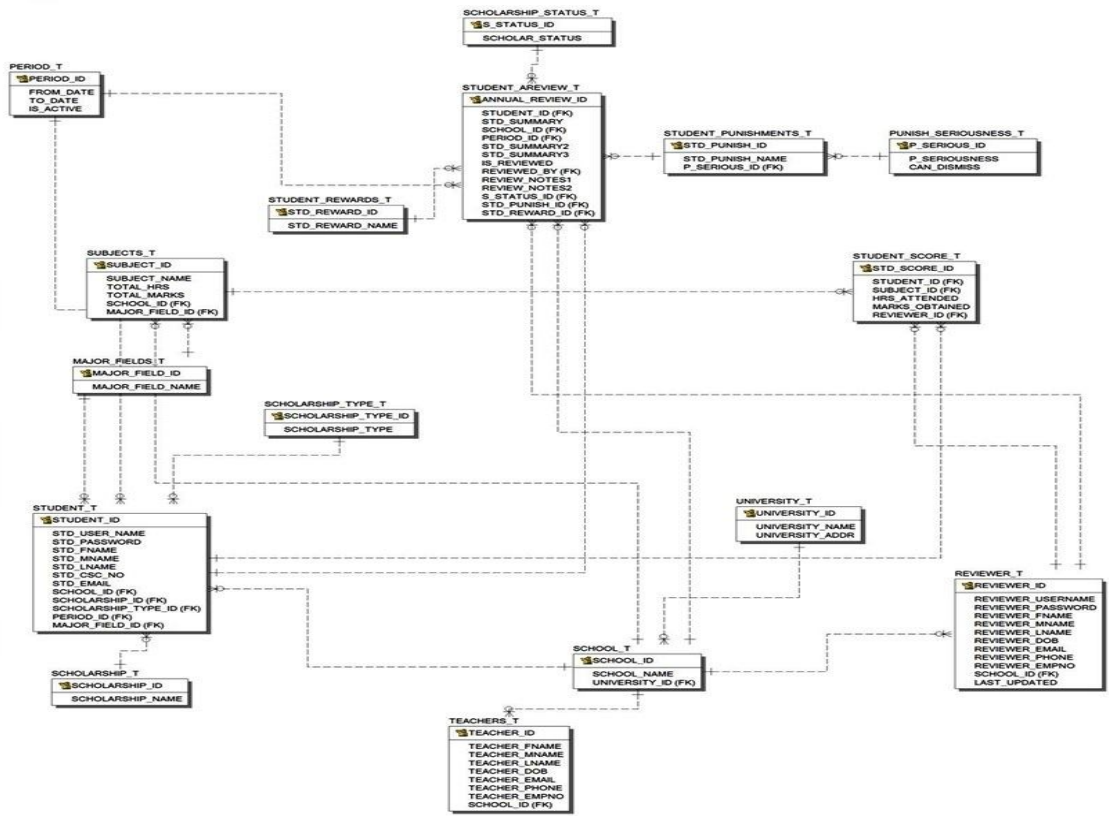

## V. SARIS PERMISSIONS/PRIVILEGES

It is assumed that the login credential information of CSC employees exists and the student can login with the login credentials provided while registering for online application.

| ACTIVITY | STUDENT | REVIEWER | CSC |
|---|---|---|---|
| Register | | √ | |
| Login | √ | √ | √ |
| Submit Annual Review | √ | | |
| View Submitted A. Review | √ | √ | √ |
| Edit Submitted A. Review | | √ | √ |
| Verify/Authenticate A. Review | | √ | |
| Submit/Edit Academic Scores | | √ | |
| View Academic Scores | √ | √ | √ |
| Submit/Edit Punishments | | √ | |
| View Punishments | √ | √ | √ |
| Submit/Edit Rewards | | √ | |
| View Rewards | √ | √ | √ |
| View Scholarship Status | √ | √ | √ |
| Edit Scholarship Status | | | √ |

Table 5.1: SARIS Access rights

## VI. PREDICTING STUDENT SUCCESS

We have gathered the information from international students living at dormitories of Chongqing University, P.R China as the dataset for our study. However, the dataset (csv) can be created using the SARIS. The table shows dataset generated using SARIS.

| STUDENT_ID | SUBJECT_FAILED | DISMISSAL_PUNISH | REWARDS | SUCCESS |
|---|---|---|---|---|
| 100121 | 2 | 0 | 0 | NO |
| 100213 | 0 | 1 | 2 | NO |
| 200128 | 5 | 0 | 2 | NO |
| 201324 | 0 | 0 | 0 | YES |
| 201217 | 1 | 0 | 0 | YES |

Table 6.1: SARIS Dataset

We have chosen the csv file as the dataset for prediction. In contrast, the other way is to make direct connection with the database for Student Success Prediction. Our method is using Weka (open source tool) for data mining purpose. The model below illustrates the proposed method for prediction:

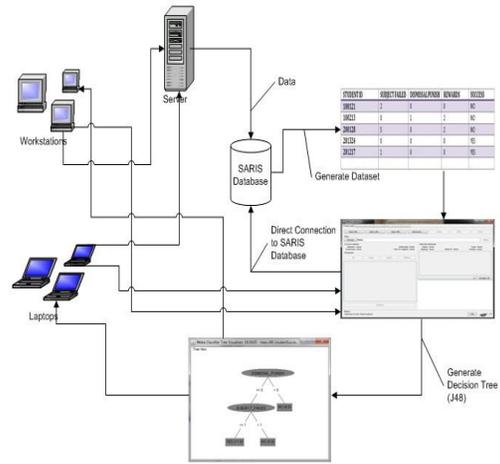

Figure 6.1: Prediction Model

## VII. CONCLUSION

We have proposed Student Annual Information System (SARIS) and a prediction method to predict students' success. SARIS manages the information of students that is required by school and the CSC. J48 decision tree concludes the student success prediction that is useful for the CSC. The proposed system can be used by any scholarship agency and any university which conducts students' annual review.